\documentclass[11pt,twoside]{article}

\usepackage{asp2006}
\usepackage{epsf}
\usepackage{graphicx}
\usepackage{lscape}
\usepackage{amsbsy}
\usepackage{amssymb}

\begin{document}
\title{Magnetic Doppler Imaging of Active Stars}

\author{O. Kochukhov and N. Piskunov}   

\affil{Department of Astronomy and Space Physics, Uppsala University, 
SE-751 20, Uppsala, Sweden}    

\begin{abstract} 
We present a new
implementation of the magnetic Doppler imaging technique, which aims at self-consistent
temperature and magnetic mapping of the surface structures in cool active stars. 
Our magnetic imaging
procedure is unique in its capability to model individual spectral features in
all four Stokes parameters. We discuss performance and intrinsic limitations of
the new magnetic Doppler imaging method. A special emphasis is given to the
simultaneous modelling of the magnetically sensitive lines in the optical and
infrared regions and to combining information from both atomic and molecular
spectral features. These two techniques may, for the first time, give us a tool
to study magnetic fields in the starspot interiors.
\end{abstract}

\section{Introduction}   

Rotational modulation of the intensity and polarization spectra of active stars offers a unique
possibility to reconstruct 2-D maps of the photospheric magnetic fields and temperature spots,
thus providing a key constraint for the theoretical modelling of the stellar activity and
dynamo. Until now \textit{Zeeman Doppler  imaging} of the active late-type stars
\citep{BDR91,HDC00,DCS03} was based upon a non-simultaneous and greatly simplified interpretation of
the average (least-squares deconvolved) Stokes $I$ and Stokes $V$ spectra. In contrast,
detailed and physically consistent methods of the stellar magnetic field mapping have been
developed \citep{PK02,KP02} and successfully applied to the high-resolution circular
and linear polarization observations of the early-type magnetic
stars \citep{KPI02,KBW04}.

Improvements in the observational techniques and instrumentation have recently made it 
possible to detect polarization signatures inside individual atomic \citep{P06} and molecular
\citep{B06} lines of the late-type stars. With these observational data in mind, we have
developed a new \textit{Magnetic Doppler imaging}  (MDI) code {\tt Invers13}. This novel tool
aims at self-consistent reconstruction of the temperature and magnetic structures in cool
active stars. A special property of our magnetic DI procedure is its capability to
realistically model individual atomic and molecular lines in all four Stokes parameters
simultaneously. Here we describe physical foundations and numerical implementation of our MDI
technique. We also present numerical experiments designed to evaluate performance of {\tt
Invers13}.

\section{Magnetic Doppler imaging with {\tt Invers13}}

The new MDI code reconstructs iteratively two-dimensional distributions of temperature and
three components of the magnetic field vector using high-resolution spectropolarimetric time
series recorded in the Stokes $I$ and $V$ or all four Stokes parameters. Inversions employ the
Levenberg-Marquardt minimization procedure and are constrained with the Tikhonov regularization
\citep{PK02}.

Our modelling of the magnetic field and cool spots is self-consistent, i.e. the local intensity
and polarization profiles are computed taking into account the Zeeman effect and the
presence of temperature inhomogeneities at the same time. 
This allows us to deduce correct field strength inside cool spots. In contrast,
all previous ZDI methods, that use only the Stokes $V$ for magnetic mapping, significantly
underestimate magnetic flux in cool spots or fail to detect it altogether.

The magnetic spectrum synthesis is carried out using realistic model atmosphere grids of cool
stars (ATLAS9, Phoenix, MARCS). The atomic and molecular concentrations are calculated with the
advanced equation of state solver \citep{VPJ98}. The local Stokes $IQUV$ parameters are computed by
solving the polarized radiative transfer equation with the quadratic DELO method 
\citep{STR00,PK02}.

\section{Numerical simulations}

Performance of the new MDI code is verified with the numerical experiments. In these tests {\tt
Invers13} infers temperature and magnetic maps from the simulated observational  data in two or
four Stokes parameters. For all tests presented here we used the stellar parameters $T_{\rm
eff}=5000$~K, $\log g=4.0$, $v\sin i=40$~km\,s$^{-1}$, $i=50^{\rm o}$. We assumed the
temperature contrast of 1500 and 750~K for the umbral and penumbral starspot regions,
respectively. The magnetic field strength was set to 2~kG in the center of a spot, reducing
to 1~kG in the penumbra. Simulations were performed for the three optical magnetically sensitive
Fe~{\sc i} lines at $\lambda$~5497.5, 5501.5, 5506.8~\AA, two IR lines Fe~{\sc i} 15648.5,
15652.9~\AA, and the TiO bandhead at 7055~\AA.

\begin{figure*}[!t]
\plotone{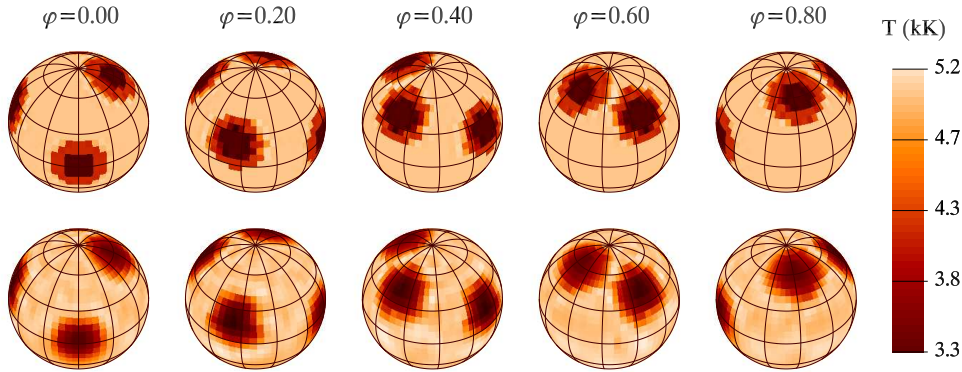}
\caption{Reconstruction of the temperature inhomogeneities with {\tt Invers13}.
\textit{Upper row:} the input temperature distribution.
\textit{Lower row:} reconstructed map.}
\label{fig1}
\end{figure*}

\begin{figure*}[!th]
\plotone{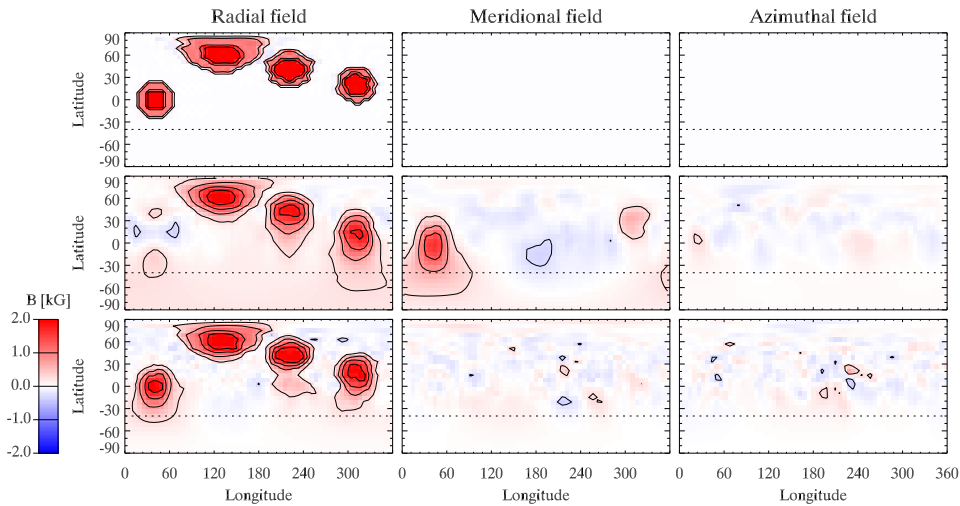}
\caption{Reconstruction of the magnetic field topology consisting of four spots 
with radial field. \textit{Upper row:} the true distribution of magnetic field.
\textit{Middle row:} reconstruction using the Stokes $IV$ spectra.
\textit{Lower row:} reconstruction using all four Stokes parameters. 
}
\label{fig2}
\end{figure*}

In the first simulation described here we studied the capability of {\tt Invers13} to infer
temperature contrast of cool starspots. Fig.~\ref{fig1} shows reconstruction of the temperature
distribution consisting of four spots located at different latitudes. For these large-scale
features the spot-to-photosphere temperature contrast is underestimated by no more than
200--300~K even when we use only the Fe~{\sc i} lines around 5000~\AA. The temperature
reconstruction is improved (errors reduced to 50--100~K) if the TiO band or the IR lines are included in the inversion.

In the second series of tests we have performed magnetic inversions for large, purely magnetic
spots containing radial, meridional and azimuthal field. Inversion for the case of radial
magnetic topology is illustrated in Fig.~\ref{fig2}. It is evident that the radial field is
successfully recovered for the high- and intermediate-latitude spots. At the same time, a strong
radial to meridional field crosstalk occurs for the low-latitude spots in the Stokes $IV$
imaging. This problem is alleviated when linear polarization data are included in the
magnetic inversion.

\begin{figure*}[!t]
\plotone{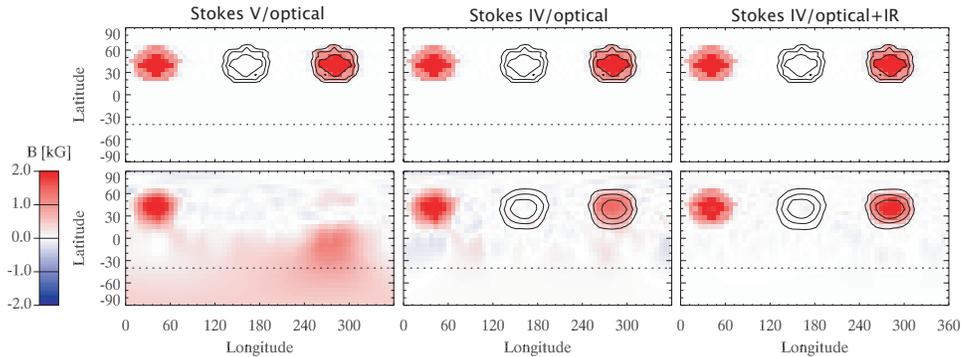}
\caption{Reconstruction of the magnetic field inside and outside cool starspots.
The 2-D images show distribution of the radial magnetic component, whereas the 
contour lines show location of cool spots.
The spot at the longitude 280$^{\rm o}$ combines reduced temperature and a
strong magnetic field. The other two spots are magnetic or
temperature only. Different rectangular plots show the true maps of surface
inhomogeneities (\textit{upper row}) and the reconstructions (\textit{lower row}) using Stokes
$V$ alone (\textit{left column}), self-consistent Stokes $IV$ mapping (\textit{middle column}) with
the optical lines, and a simultaneous $IV$ mapping with the optical and IR lines (\textit{right column}).
}
\label{fig3}
\end{figure*}

Finally, we have tested possibility to probe magnetic fields inside cool starspots. In
Fig.~\ref{fig3} we illustrate the adopted structure of the temperature and magnetic
inhomogeneities and show results of different magnetic inversions. We find that if the magnetic
field is studied using the Stokes $V$ alone, without accounting for the temperature
inhomogeneities (as it is done in all previous implementations of ZDI), the resulting map
misses entirely the magnetic flux in the regions of
reduced temperature. On the other hand, our standard MDI reconstruction based on the
self-consistent interpretation of the Stokes $I$ and $V$ spectra recovers correctly the
geometry of the field, although the field strength is underestimated by about 50\% in the
starspot center. The agreement between the true and reconstructed maps can be improved
significantly if the IR Fe~{\sc i} lines are incorporated in the inversion. This becomes
possible due to a much smaller spot-to-photosphere flux contrast in the IR and the resulting
larger relative contribution of the spot polarization signatures to  the disk-integrated Stokes
$V$ spectra. The polarization diagnostic of the TiO molecular bands allows us to achieve similar
improvement in the magnetic mapping of the starspot regions, although the MDI with molecular
lines suffers from reduced spatial resolution due to a large width of molecular features.
However, the main advantage of the simultaneous analysis of molecular and atomic  lines is the
possibility to use the same observational material for the inversion, whereas costly multisite
campaigns are needed to secure simultaneous visual and IR high-resolution spectra.

\end{document}